# Science from the Moon: The NASA/NLSI Lunar University Network for Astrophysics Research (LUNAR)

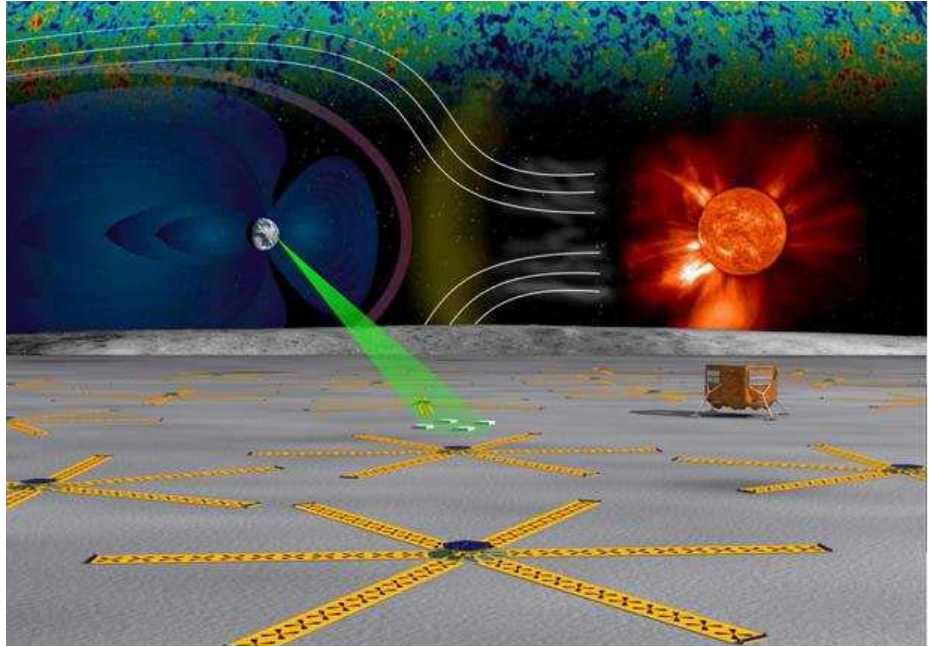

## The LUNAR Consortium:

*Jack Burns[1], E. Hallman, D. Duncan, J. Darling, & J. Stocke*, University of Colorado at Boulder
*J. Lazio & K. Weiler*, Naval Research Laboratory
*J. Hewitt*, Massachusetts Institute of Technology
*C. Carilli, R. Bradley, T. Bastian & J. Ulvestad*, National Radio Astronomy Observatory
*J. Kasper & L. Greenhill*, Smithsonian Astrophysical Observatory
*R. MacDowall, S. Merkowitz, J. McGarry, Zagwodzki, P. Yeh, H. Thronson, & S. Ne*, NASA Goddard Space Flight Center
*D. Currie*, University of Maryland at College Park
*T. Murphy*, University of California at San Diego
*S. Furlanetto & A. Mesinger*, University of California at Los Angeles
*A. Loeb, J. Pritchard & E. Visbal*, Harvard University
*D. Jones*, Jet Propulsion Laboratory
*G. Taylor*, University of New Mexico
*K. Nordtvedt*, Northwest Analysis
*J. Bowman*, California Institute of Technology
*J. Grunsfeld*, NASA Johnson Space Center
*S. Bale*, University of California at Berkeley
*B. Wandelt*, University of Illinois at Urbana-Champaign
*H. Falcke*, Radboud University, Nijmegen & ASTRON, Dwingeloo, Netherlands

[1]Corresponding author. Center for Astrophysics & Space Astronomy, 389 UCB, University of Colorado, Boulder, CO 80309, jack.burns@cu.edu, 303-735-0963, http://lunar.colorado.edu.

A White Paper submitted to the Planetary Sciences Decadal Survey

# Science from the Moon: The NASA/NLSI Lunar University Network for Astrophysics Research (LUNAR)

**Abstract.** The Moon is a unique platform for fundamental astrophysical measurements of gravitation, the Sun, and the Universe. Lacking a permanent ionosphere and, on the farside, shielded from terrestrial radio emissions, a radio telescope on the Moon will be an unparalleled heliospheric and astrophysical observatory. Crucial stages in particle acceleration near the Sun can be imaged and tracked. The evolution of the Universe before and during the formation of the first stars will be traced, yielding high precision cosmological constraints. Lunar Laser Ranging of the Earth-Moon distance provides extremely high precision constraints on General Relativity and alternative models of gravity, and also reveals details about the interior structure of the Moon. With the aim of providing additional perspective on the Moon as a scientific platform, this white paper describes key research projects in these areas of astrophysics from the Moon that are being undertaken by the NLSI-funded LUNAR consortium.

The NASA Lunar Science Institute (NLSI) recently funded 7 mostly university-based teams to study science of, on, and from the Moon. The LUNAR consortium was selected by the NLSI for astrophysical research and education that focuses on the key, unique instruments that most effectively take scientific advantage of sites on the lunar surface – low frequency heliophysics and cosmology, and lunar laser ranging. We are submitting this white paper to the Planetary Sciences Decadal Survey to provide additional perspective on the value of Moon for conducting cutting-edge research in astrophysics and gravitational physics by describing our key projects for LUNAR. This program of astrophysics from the Moon complements as well as takes advantage of expected scientific infrastructure on the Moon during the next few decades.

1. **Lunar Radio Array for Cosmology & Astrophysics**

The Dark Ages represent a new frontier in cosmology, the era between the genesis of the cosmic microwave background (CMB) at recombination (when electrons and protons combine as hydrogen) and the formation of the first stars (Figure 1). During the Dark Ages, before the Universe was lit by the first stars, the entire baryonic content of the Universe was neutral hydrogen (HI) and helium in a diffuse intergalactic medium (IGM), and the only means known to study this medium is through the redshifted HI 21-cm line [1]. This HI signal represents potentially a very rich cosmological dataset — for a portion of the Dark Ages, the physics is sufficiently simple that the HI signal can be used to constrain fundamental cosmological parameters in a manner similar to that of CMB observations, but the spectral nature of the signal allows the evolution of the Universe as a function of redshift ($z$) to be followed.

During the Dark Ages, overdensities of dark matter gravitationally collapse, pulling with them HI. These overdensities eventually produce the first stars and cosmic twilight of the Universe probably at $z \sim 30$ (about 100 million years after the Big Bang). These first stars go through their stellar lifecycle quickly, exploding as Type II supernovae and creating large bubbles of ionized gas. As more of these first stars form, they produce additional bubbles which begin to merge and form a fully ionized IGM. During this period, the first black holes and quasars are thought to form, generating intense X-ray emission and further driving the reionization. The Epoch of Reionization (EoR) likely ends at $z > 6$–10 (about 500 million years after the Big Bang). The details of this schematic picture at present remain an active area of research and *with exceptionally limited observational confirmation*. The Moon is likely the only site in the inner solar system for Dark Ages observations as significant obstacles exist to ground-based telescopes, including heavy use of the relevant portion of the spectrum by both civil and military transmitters, distortions introduced by the Earth's upper atmosphere (ionosphere), and solar radio emissions.

Recently, the global or sky-averaged HI signal expected from the end of the Dark Ages through the EoR ($6 < z < 300$) has been modeled [2]. That model spectrum is shown on the right side of Fig. 1. After recombination, the gas kinetic temperature ($T_K$) (and spin temperature, $T_S$) tracks the CMB temperature ($T_{CMB} \sim (1+z)$) down to $z \approx 300$ and then declines below it ($T_K \sim (1+z)^2$) producing the right-most absorption trough in Figure 1, corresponding to $5 < z < 50$ MHz during the Dark Ages. The spin temperature of the 21-cm transition closely follows $T_K$ through collisional coupling during this time. As the Universe expands and the density decreases, collisional coupling is not effective so the spin temperature again tracks the CMB via radiative coupling. Thus, the HI absorption trough closes at $z \approx 20$-30. Any exotic heating mechanisms, such as Dark Matter decay, will be evident in the shape and peak of this Dark Ages absorption profile [1].

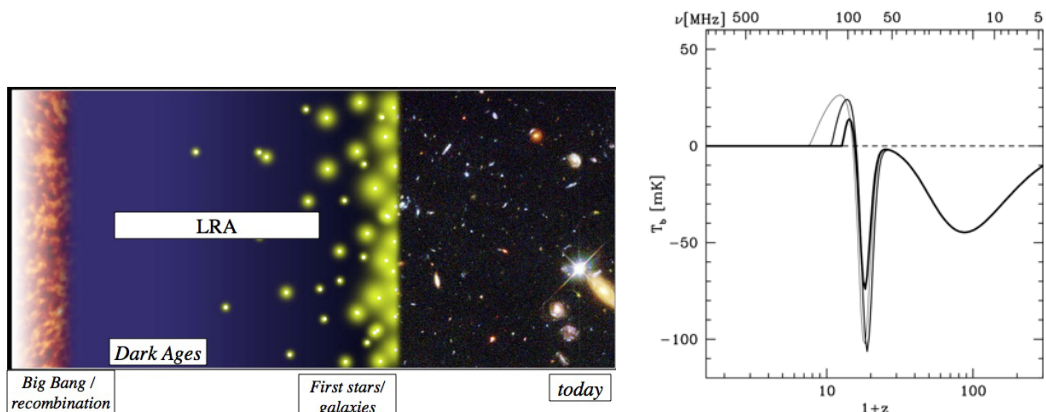

**FIGURE 1.** (Left) Schematic illustration of the evolution of the Universe (Djorgovski et al. & Digital Media Center, Caltech). The first stars and black holes form during the end of the Dark Ages transforming the IGM from a neutral (HI) to an ionized state. LRA is the Lunar Radio Array that would observe highly redshifted 21-cm signals that will constrain the first epoch of feedback. (Right) Evolution of the global (all-sky) redshifted 21-cm brightness temperature as function of redshift (and frequency) for 3 models of the first galaxies [2].

A second, narrower but deeper, absorption trough is centered at $z \approx 75$ MHz ($z = 15$-20). This HI absorption is produced after the first stars turn on and begin to flood the IGM with Ly$\alpha$ UV photons. This produces a Ly$\alpha$ resonance with the HI hyperfine transition creating enhanced absorption via the Wouthuysen-Field effect [3]. The shape, width, and center frequency of this trough depend upon the astrophysical details of the first galaxies including the nature of the first stars and X-ray emission from the first black hole-driven active galactic nuclei (AGNs).

As the IGM is heated, most likely by feedback via soft X-rays from the first AGNs, the kinetic and spin temperatures rise above $T_{CMB}$, producing redshifted 21-cm emission.

The brightness temperature depends critically upon the neutral fraction which diminishes as the IGM becomes further ionized. This emission occurs at $z \approx$ 100-200 MHz, near the end of the EoR ($z<12$). Ultimately, reionization kills off this redshifted 21-cm signal. Observations of the global HI emission at these frequencies can provide important constraints on the formation and nature of the first black holes.

Highly redshifted 21-cm signals from the Dark Ages and the EoR have the potential of producing a very rich cosmological database rivaling that of the CMB. Constraints on the properties of very high-$z$ galaxies and their effects on the surrounding IGM, Dark Energy, Dark Matter decay, and the nature of the field that drove the Universe during Inflation may emerge from this database [1]. Furthermore, the database will be *three dimensional*, permitting us to probe the evolution of HI and feedback via redshifted 21-cm tomography.

Efforts are now underway to build the first generation of EoR experiments on the ground to probe the low redshift, higher frequency portion of the spectrum in Fig. 1 [4]. A key challenge for all of these experiments will be Galactic and extragalactic foregrounds. This next decade will reveal if these telescopes can accurately remove foregrounds at the levels needed to see the redshifted 21-cm EoR signal. This could lead to the development of a low frequency Square-Kilometer Array (SKA) [5]. There is likely to be a limit to ground-based systems, whereas the farside of the Moon has significant advantages as it effectively lacks an ionosphere, has a demonstrated radio-quiet environment free of interference [6], and is shielded from powerful solar emissions during the lunar night. In order to probe the beginning of the EoR and into the Dark Ages, the lunar farside is likely the only location in the inner solar system from which such observations will be successful.

Even before a large, farside lunar array, a single dipole antenna aboard a spacecraft in lunar orbit or on the lunar surface operating in the frequency regime of ≈ 25–75 MHz has much potential. Such pathfinding observations hold the promise of testing models of the expected average all-sky absorption signals as illustrated in Fig. 1. These data would be the first to sample the formation of the first stars and black holes at z ≈ 10–20, thus providing concrete measurements on how the Universe was reionized during this first generation of feedback. A stable broadband spectrometer with carefully controlled noise on the spacecraft bus will be required to carry out high dynamic range observations as will innovative, robust algorithms for removal of the significant foregrounds.

A next step will be a modest-size interferometer of ~100 elements deployed on the lunar farside. Such an array could verify and extend to lower frequencies the results on the EoR from ground-based low frequency telescopes. Such an interferometer could be deployed robotically with astronaut supervision. This precursor telescope would then evolve into a more capable ~$10^4$ element array for imaging tomography of the Dark Ages and the beginning of the EoR [7,8]. The deployment of such a Lunar Radio Array will be facilitated by a next generation of heavy-launch vehicle enabled by the exploration architecture.

## 2. Radio Heliophysics from the Moon

Fundamental understanding of particle acceleration within the heliosphere could result from a simple low-frequency radio array on the nearside of the Moon. The Radio Observatory for Lunar Sortie Science (ROLSS) is an experiment that could be deployed robotically or by astronauts during a sortie landing. ROLSS would image emission produced by accelerated electrons in the solar corona and inner heliosphere, provide advance warning at the Moon of radiation events, conduct pathfinding astrophysical imaging of the sky, and serve as a pathfinder for the Cosmology LRA.

High energy particle acceleration occurs in diverse astrophysical environments including the Sun and other stars, supernovae, black holes, and quasars. Fundamental problems include understanding the mechanisms and sites of this acceleration, in particular the roles of shock waves and magnetic reconnection. Within the inner heliosphere, an interval of 1–10 solar radii ($R_s$) from the Sun, solar flares and coronal mass ejections (CMEs) are efficient particle accelerators. Low frequency observations are an excellent remote diagnostic because electrons accelerated by these structures can produce intense radio bursts. The intensity of these bursts makes them easy to detect, but they also provide information about the acceleration regions. For example, the radio burst mechanisms discussed here involve emission at the local plasma frequency, $f_p \approx 9 n_e^{1/2}$ kHz, or its harmonics, where $n_e$ is the electron density in cm$^{-3}$. With a model for $n_e$, $f_p$ can be converted into a height above the corona, and the gradient of $f_p$ can be converted into radial speed.

Solar radio bursts are one of the primary remote signatures of electron acceleration in the inner heliosphere and our focus is on two emission processes, referred to as Type-II and Type-III radio bursts. Type II bursts originate from suprathermal electrons (E>100 eV) produced at shocks. These shocks are generally produced by CMEs as they expand into the heliosphere with Mach numbers greater than unity. Emission from a Type-II burst drops slowly in frequency as the shock moves away from the Sun into lower density regions at speeds of 400 - 2000 km s$^{-1}$. Type III bursts are generated by fast (2 - 20 keV) electrons from magnetic reconnection. As the fast electrons escape at a significant fraction of the speed of light into the heliosphere along magnetic field lines, they produce emission that rapidly drops in frequency. Densities in the critical range of the solar corona and inner heliosphere where a strong CME may develop and accelerate electrons corresponds to frequencies below tens of MHz which are difficult if not impossible to image on Earth due to ionospheric distortion and absorption. These measurements must therefore be made from space.

The Radio Observatory for Lunar Sortie Science (ROLSS) is a concept for a low-frequency array that would be deployed during or before the first lunar sorties. ROLSS is designed to conduct radio imaging observations of the Sun to address the particle acceleration questions discussed above. The concept of low-frequency radio imaging from space is not new, but has generally been based on a constellation of free-flying spacecraft, each with an antenna [8,9]. Building an array on the Moon has three advantages over the spacecraft. First, with the lunar surface as a structural support, the antennas can be lighter. Second, fuel is not spent to maintain the array configuration. Finally, the Moon blocks half the sky, simplifying the imaging.

ROLSS uses three arms arranged in a Y configuration, much like the Very Large Array (VLA). Instead of VLA dishes, ROLSS would use simple electrically-short dipole antennas. Each arm is a thin, 500-m long polyimide film (PF) with metal circuit traces for the antennas and to bring the signals back to a central station (CS). The CS would house signal processing, solar panels, batteries and heaters, and an Earth-directed antenna.

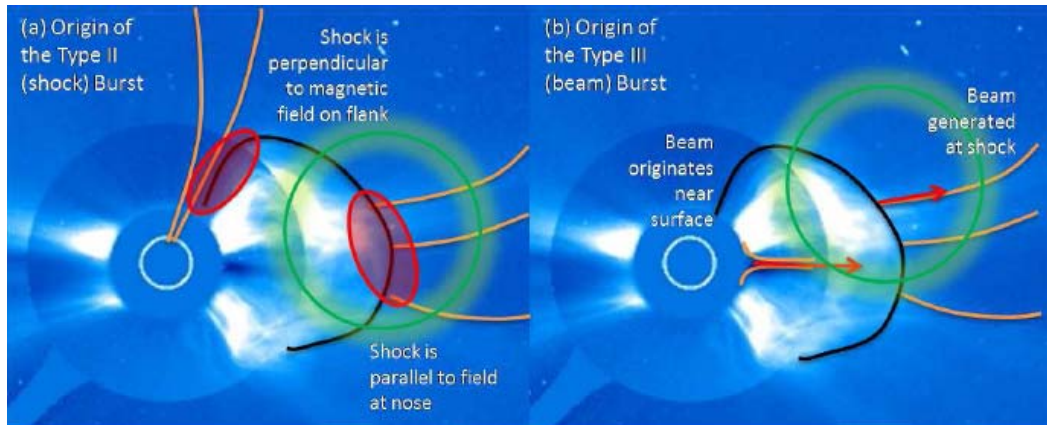

**FIGURE 2:** (a) Possible source regions of Type-II bursts created by electrons accelerated near shock surfaces. (b) Possible source regions of Type-III bursts due to electron beams escaping along magnetic field lines into the heliosphere. In both images are of a CME observed by SOHO and superposed lines indicate notional magnetic field lines (orange), shock waves (black), acceleration sites (red), and ROLSS angular resolution at 10 MHz (green).

3. **Gravitational Physics & Lunar Structure via Lunar Laser Ranging**

Gravity is the force that holds the universe together, yet a theory that unifies it with other areas of physics still eludes us. Testing the very foundation of gravitational theories, like Einstein's theory of general relativity, is critical in understanding the nature of gravity and how it relates to the rest of the physical world. Lunar laser ranging (LLR) has been the workhorse for testing general relativity over the past 40 years. The three retroreflector arrays put on the moon by the Apollo astronauts and one of the Soviet Luna arrays continue to be useful targets, and have provided the most stringent tests of the Strong Equivalence Principle and the time variation of Newton's gravitational constant. The LUNAR team is focused on addressing 5 key science questions for which LLR is a major contributor:

1. Is the Equivalence Principle exact?
2. Does the strength of gravity vary with space and time?
3. Do extra dimensions or other new physics alter the inverse square law?
4. What is the nature of spacetime?
5. What is the interior structure and composition of the Moon?

These questions are discussed in detail in two other white papers [10,11]; here we briefly describe some the activities of the LUNAR team that are aimed at making sure the next generation of LLR is capable of answering these questions.

The first LLR measurements had a precision of about 20 cm. Since 1969, multiple stations have successfully ranged to the lunar retroreflectors and have increased the range precision by a factor of 10 to the level of a few centimeters. Poor detection rates have historically limited LLR (not every laser pulse sent to the Moon results in a detected return photon). However, the relatively new APOLLO instrument at the Apache Point telescope has a large collecting area and uses very efficient avalanche photodiode arrays such that thousands of detections are recorded (even multiple detections per pulse) leading to a statistical uncertainty of about 1 mm [12].

The dominant random uncertainty per photon in modern LLR stations stems from the orientation of the reflector array and the associated spread of pulse return times. Additionally, at the millimeter level of precision, systematic errors associated with lunar arrays start to become significant. Going beyond this level of precision will likely require new lunar retroreflectors or laser transponders that are more thermally stable and are designed to reduce or eliminate orientation dependent pulse spreading. **Ground station technology has now advanced to a point where further improvements in range precision, and consequently the tests of general relativity, will be limited by errors associated with the lunar arrays.**

Significant advances in lunar laser ranging will require placing modern retroreflectors and/or active laser ranging systems at new locations on the lunar surface. Ranging to new locations will enable better measurements of the lunar librations, aiding in our understanding of the interior structure of the moon. More precise range measurements will allow us to study effects that are too small to be observed by the current capabilities as well as enabling more stringent and crucial tests of gravity.

Large single cube corners (>10cm) can potentially be made to provide similar return rates as the Apollo arrays without significant pulse spreading. Several cubes could be deployed around a landing site with enough separation that responses do not overlap when seen by the earth stations to further increase the total response. The LUNAR LLR team is looking at both traditional solid cubes and modern hollow cubes as potential solutions. In addition, active laser ranging systems, such as asynchronous laser transponders and laser communication terminals, are also potential options that additionally have applications for Mars and other interplanetary ranging that have science goals similar to those of lunar ranging.

The analysis of LLR data requires a sophisticated model of the solar system ephemeris that also includes all the significant effects that contribute to the range between the Earth station and the lunar retroreflector. These models typically include hundreds of parameters that describe solar system bodies and effects such as the relativistic corrections, tidal distortions, plate motion, and atmospheric propagation delay.

A number of models have been developed over the past 40 years to analyze LLR data. The model developed at JPL was recently used to produce the best limits on the Strong Equivalence Principle violation and time variation of the gravitational constant [13]. The open-source Planetary Ephemeris Program (PEP) is undergoing a major upgrade for LLR analysis at the Harvard-Smithsonian Center for Astrophysics (CfA). It was used for one of the first LLR analysis to test the Strong Equivalence Principle [14] and was recently used to test for Lorentz violation [15].

To take advantage of the next generation of LLR instruments (and today's millimeter level data), these codes need to be modified and rigorous theoretical work needs to be performed to permit tests of new ideas in physics. Substantial effort is also required to address the multitude of effects that will contribute at the sub-millimeter level. Many of these effects will be scientifically interesting in their own right. The LUNAR LLR team is laying out the groundwork for the development of the next generation of LLR analysis tools.

Finally, the LLR team is working with the rest of the LUNAR collaboration on issues common to the different projects. These include emplacement of arrays of detectors over large areas of the lunar surface and innovative deployment strategies using humans and/or robotic rovers. These deployment studies will not only be of benefit to the LUNAR collaboration, but also to the other projects that require instruments at multiple sites.